\documentclass[preprint,aps,showpacs]{revtex4}

\usepackage{graphicx}

\begin{document}

\title{Interfacial phase transitions in twinning-plane superconductors}
\author{F. Clarysse}
\thanks{Present address: Department of Mathematics, 
Heriot-Watt University, Edinburgh
EH14 4AS, United Kingdom.}
\author{J.O. Indekeu}
\affiliation{Laboratorium voor Vaste-Stoffysica en Magnetisme,
Katholieke Universiteit Leuven, B-3001 Leuven, Belgium.}

\date{\today}

\begin{abstract}

Within the framework of Ginzburg-Landau theory we study the rich variety of 
interfacial phase transitions in twinning-plane superconductors. We show that
the phase behaviour strongly depends on the transparency of the twinning plane
for electrons measured by means of the coupling parameter $\alpha_{{\rm TP}}$. By analyzing the solutions of the Ginzburg-Landau equations in the limit of perfectly transparent twinning planes, we predict a first-order interface delocalization transition for all 
type-I materials. We further perform a detailed study of the other limit in
which the twinning plane is opaque. The phase diagram proves to be very rich and fundamentally different from the transparent case, recovering many of the
results for a system with an external surface. In particular both first-order
and critical delocalization transitions are found to be possible, accompanied by
a first-order depinning transition. We provide a comparison with experimental 
results and discuss the relevance of our findings for type-II materials.

\end{abstract}


\maketitle


\section{Introduction}
\label{sec:intro}

In recent years, local enhancement of superconductivity has been predicted to
provide the mechanism to induce several intriguing interfacial phase transitions
in type-I superconductors \cite{IND,IND1,CJB,BAC,MON}. Typical phase diagrams are
calculated using the Ginzburg-Landau (GL) theory in which the enhancement
is accounted for by allowing the extrapolation length $b$ to be negative. The
microscopic origin of this parameter remains an unsolved problem making
the experimental verification of the theoretical results non-trivial. So far, 
the most feasible realization of a negative extrapolation length seems to 
originate from the concept of twinning-plane superconductivity (TPS), a well 
understood phenomenon that occurs, e.g., in Sn, In, Nb, Re and Tl \cite{BUZ}. A 
twinning plane (TP) is a defect plane representing the boundary between two 
single-crystal regions or twins and, consequently, the physics encountered in 
the behaviour of a superconducting/normal interface near TP's is the natural 
analogy to grain-boundary wetting or interface depinning, a topic which has been
well studied in magnetic systems \cite{ABR,SEV,IGL}.

The characteristic feature of the original GL approach of TPS is the a priori
assumption that the TP is perfectly transparent for electrons at the microscopic
level which implies that the superconducting order parameter $\psi$ is 
continuous at the TP \cite{BUZ}. Subsequent extensions of the theory relax this
assumption allowing a discontinuity in $\psi$ \cite{AND,GES,MIN,SAM}. More
specifically, a second phenomenological parameter, $\alpha_{{\rm TP}}$, is 
introduced to describe the coupling between the twins such that, by means of 
$\alpha_{{\rm TP}}$,
one can mimic the effect of microscopically tuning the TP from completely
transparent to completely opaque for electrons. In this paper we present an 
overview of the variety of interfacial phase transitions in the two limiting 
cases to develop a thorough understanding of the influence of the transparency.

Earlier work~\cite{BAC} has focused on the case of mixed bulk boundary 
conditions with the bulk normal (N) phase on one side and the bulk 
superconducting (SC) phase on the other side of the TP, at bulk two-phase 
coexistence. This is appropriate for the study of the proper depinning 
transition of an interface that is initially pinned at the TP. Here we choose to 
settle for the configuration of equal bulk conditions, that is, we impose the 
bulk N phase on both sides of the TP. In so doing we are no longer restricted to
the case of bulk two-phase coexistence and this allows us to establish the 
complete magnetic field versus temperature phase diagram for a given material. 
This type of diagram is accessible to experimental verification and is relevant 
for comparing the present results with known TPS phase 
diagrams~\cite{BUZ,MIN,SAM}.

The solutions of the GL equations strongly depend on the boundary conditions
imposed at the TP itself which in turn relates to the level of transparency, 
i.e. the value of $\alpha_{{\rm TP}}$. For highly transparent planes, corresponding to 
the limit $\alpha_{{\rm TP}} \rightarrow 0$, it is natural to consider fully symmetric
profiles for the order parameter. In the opposite limit, $\alpha_{{\rm TP}} 
\rightarrow \infty$, the TP is completely opaque for electrons and both sides
are largely independent. In this case there is a wide range of possible
solutions, including profiles with $\psi$ identically zero at one side of the
TP. The latter are refered to as \emph{wall} solutions, since they
are equivalent to the ones found in a type-I superconductor with an external
surface or wall characterized by a negative extrapolation length 
$b$~\cite{IND,IND1}.
Therefore we anticipate that, in the opaque limit, we will recover to a great
extent the results of a wall system. This is very different from the case of
complete transparency, for which drastic qualitative modifications are predicted
compared to the case with a wall.

The outline of the paper is as follows. In the next section we collect the main 
ideas of the GL theory applied to twinning-plane superconductors.
Section~\ref{sec:transp} covers the results for perfectly transparent TP's. We
calculate in detail the phase diagrams and provide a comparison with the
predicted TPS diagrams as described in Ref.~\cite{BUZ}. The fully opaque system 
is the subject of Section~\ref{sec:opaque}. We present a classification of the 
various solutions and establish their stability to derive the phase behaviour.
We summarize our main results and discuss the experimental relevance in 
Section~\ref{sec:conc}. 

\section{Ginzburg-Landau theory for twinning-plane superconductors}
\label{sec:gl}

We consider a type-I superconductor with a TP located at $x=0$ and impose on 
both sides the N phase, with $\psi =0$, as the bulk condition. The GL 
free-energy functional has the form
\begin{equation}
\Gamma[\psi,{\bf A}]=\int_{-\infty}^{+\infty}{\cal G}[\psi,{\bf A}]{\rm d}x+ \Gamma_{\rm TP}(\psi_-,\psi_+), \label{eq:gammatp1}
\end{equation}
with the free-energy density ${\cal G}$ given by
\begin{equation}
{\cal G}= \epsilon|\psi|^2+\frac{\beta}{2}|\psi|^4+\frac{1}{2m}\left| \left(
\frac{\hbar}{i}{\bf \nabla}-2e{\bf A} \right) \psi\right|^2+\frac{[{\bf
\nabla}\times{\bf A} - \mu_0{\bf H}]^2}{2\mu_0}. \label{eq:enedens}
\end{equation}
As usual, $\epsilon \propto T-T_c$, where $T_c$ is the bulk critical temperature
which must be distinguished from the second critical temperature in the system,
$T_{{\rm c, TP}}$, below which local superconductivity sets in at the TP in zero magnetic field.
Since $T_{{\rm c,TP}}$ was experimentally \cite{BUZ} proved to be only slightly 
higher than $T_c$, the use of the GL theory is justified. Further, $\beta >0$ is
a stabilizing parameter and ${\bf A}$ is the vector potential. We choose the 
applied magnetic field ${\bf H}={\rm H}{\bf e}_z$ parallel to the TP. Using the 
notation $\psi_-\equiv \psi(0^-)$ and $\psi_+ \equiv \psi(0^+)$, the local 
contribution $\Gamma_{\rm TP}$ in (\ref{eq:gammatp1}) reads
\begin{equation}
\Gamma_{\rm TP}(\psi_-,\psi_+)=\frac{\hbar^2}{2mb}(|\psi_+|^2+|\psi_-|^2)+
\frac{\hbar^2}{2m\alpha_{{\rm TP}}}\left|\psi_+ - \psi_-\right|^2. 
\label{eq:localenetp}
\end{equation}
The first term, with $b<0$, describes the enhancement of superconductivity and
was introduced by Khlyustikov and Buzdin~\cite{BUZ} to reproduce theoretically
the observed TPS phase diagrams. The phenomenological parameter $b$ is the
extrapolation length and can be related to the temperature difference 
$T_c-T_{{\rm c,TP}}$. In addition, we have followed others~\cite{BAC,AND,GES,MIN,SAM}
by adding a second term in (\ref{eq:localenetp}) to describe the coupling 
between the twins. In so doing, we allow the SC wave function to be 
discontinuous across the TP, hence in general $\psi_- \neq \psi_+$. The 
coupling constant $\alpha_{{\rm TP}}$ can be expressed in terms of the Fermi velocity 
and either the transmission or reflection coefficient for electrons, thus fully in 
terms of microscopic properties~\cite{GES}. We note that for $\alpha_{{\rm TP}}>
0$, 
the phase of the wave function is continuous at the TP, while for 
$\alpha_{{\rm TP}}<0$ a phase jump of $\pi$ can occur~\cite{AND}. We omit the latter 
possibility and restrict our attention to 
$\alpha_{{\rm TP}} >0$.

In what follows we assume translational invariance in the $y$- and 
$z$-directions and choose the gauge so that ${\bf A}=(0,A(x),0)$. It proves to
be convenient to adopt the rescaling introduced in earlier work~\cite{IND1} 
using the two basic length scales of the superconductor,
i.e. the zero-field coherence length $\xi$ and the magnetic penetration depth
$\lambda$ defined by
\begin{equation}
\xi^2=\frac{\hbar^2}{2m|\epsilon|} \ \ \  , \ \ \      
\lambda^2=\frac{m\beta}{\mu_0q^2|\epsilon|}. \label{eq:xilam}
\end{equation}
The ratio of $\lambda$ to $\xi$ gives the GL parameter $\kappa$, with 
$\kappa< 1/ \sqrt2$ for type-I materials. We use $\xi$ to scale the distances
but, for simplicity, retain the notation $x$ for the dimensionless coordinate
$x/\xi$ perpendicular to the TP. For the magnetic quantities $A$ and $H$ we 
introduce the dimensionless $a$ and $h$ defined by
\begin{equation}
a=\frac{2e\lambda}{\hbar}A \ \ \ ,  \ \ \ h=\frac{2e \lambda^2 \mu_0}{\hbar}H,
\label{eq:aenh}
\end{equation}
and rescale the wave function $\psi$ according to
\begin{equation}
\varphi=\psi/\psi_{eq},
\end{equation}
where $\psi_{eq}=\sqrt{|\epsilon|/\beta}$ is the equilibrium value of the SC order 
parameter for $T<T_c$. Clearly, $\varphi$ attains the value 1 in the bulk SC
phase. Finally, we rescale the free energy $\Gamma$ divided by the surface area
$S$ such that $\gamma=\Gamma\beta/(\epsilon^2\xi S)$, yielding
\begin{eqnarray}
\gamma[\varphi,a] = \int_{-\infty}^{+\infty} dx\left\{
\pm\varphi^2+\frac{\varphi^4}{2}+\dot{\varphi}^2 +
\frac{a^2\varphi^2}{\kappa^2}+(\dot a -h )^2 \right\} \nonumber \\  +
\frac{\xi}{b}(\varphi_-^2+\varphi_+^2) +
\frac{\xi}{\alpha_{{\rm TP}}}(\varphi_--\varphi_+)^2. \label{eq:gammatp2}
\end{eqnarray}
The $\pm$ refers to the sign of $T-T_c$. Minimization of $\gamma$ with respect 
to $\varphi$ and $a$ yields the well-known GL equations
\begin{equation}
\ddot{\varphi}=\pm\varphi+a^2\varphi/\kappa^2+\varphi^3, 
\label{eq:gleqtp1}
\end{equation}
and
\begin{equation}
\ddot{a}=a\varphi^2/\kappa^2 .
\label{eq:gleqtp2}
\end{equation}
In addition, two coupled boundary conditions are obtained from stationarity with respect to $\varphi_-$ and $\varphi_+$
\begin{equation}
\dot{\varphi}_-= -\frac{\xi}{b}\varphi_-
+\frac{\xi}{\alpha_{{\rm TP}}}(\varphi_+-\varphi_-), \label{eq:boundtp1}
\end{equation}
\begin{equation}
\dot{\varphi}_+= \frac{\xi}{b}\varphi_+
+\frac{\xi}{\alpha_{{\rm TP}}}(\varphi_+-\varphi_-), \label{eq:boundtp2}
\end{equation}
while the vector potential and its first derivative must be continuous at 
$x=0$,
\begin{equation}
a(x=0^-)=a(x=0^+) \ \ \ , \ \ \  \dot a(x=0^-)=\dot a(x=0^+). \label{eq:a0}
\end{equation}
In the subsequent sections we aim at solving the above equations for both the
transparent and the opaque limit.

\section{Transparent twinning planes}
\label{sec:transp}

\subsection{Boundary conditions and solutions}
\label{subsec:introtra}

For highly transparent TP's, $\alpha_{{\rm TP}} \rightarrow 0$ and we recover the original description of TPS \cite{BUZ} with a continuous order parameter at the TP, thus $\varphi_+=\varphi_-$. Consequently, it is natural to look for fully symmetric solutions for $\varphi(x)$ of the differential 
equation~(\ref{eq:gleqtp1}). In practice this is done by initially restricting 
ourselves to one half space, say $x>0$. The solution in the other half space $x<0$ can then be constructed using $\varphi(x)=\varphi(-x)$, with $\varphi(x)$ the solution for the semi-infinite system. Note that this makes the problem very
similar to the one of a semi-infinite system with an external surface~\cite{IND,IND1}, the only difference emerging from the behaviour
at $x=0$ which is manifested most clearly in the boundary condition for the 
vector potential. Indeed, in the presence of an external surface, stationarity 
of $\gamma$ with respect to $a(0)$ results in $\dot a(0)=h$. In the present 
situation, however, owing to the assumed symmetry in the profile for 
$\varphi(x)$ the vector potential $a(x)$ will be anti-symmetric with respect to 
$x$, i.e. $a(-x)=-a(x)$ and thus obeys the condition $a(0)=0$. The boundary
conditions for the wave function reduce in this limit to
\begin{equation}
\dot\varphi_+=-\dot\varphi_-=\frac{\xi}{b}\varphi_+ .
\label{eq:boundtptrans}
\end{equation}

In the following analysis, we will be interested in two types of solutions 
distinguished by their asymptotic behaviour for $|x| \rightarrow \infty$. The
enhancement of superconductivity near the TP will typically induce a SC sheath
(with the bulk of the system prepared in the N phase) and it is precisely the
thickness of this surface sheath that characterizes the solution. Clearly, if
this thickness is finite superconductivity disappears for $|x| \rightarrow 
\infty$ and the magnetic field penetrates such that
\begin{equation}
\varphi(\pm \infty)=0 \ \ \ , \ \ \  \dot a(\pm\infty)=h. 
\label{eq:bulktranstp1}
\end{equation}
Adapting the terminology of~\cite{IND,IND1} which follows from the 
analogy to wetting transitions in adsorbed fluids \cite{DIE}, this 
solution is referred to as a \emph{partial wetting} state. We remark that this 
class of solutions also includes the so-called null solution without any SC 
phase in the system (thus describing a SC sheath with zero thickness). On the 
other hand, in the case of \emph{complete wetting} the sheath can be 
macroscopically thick so that it fully separates the N phase from the TP. These 
solutions are possible only when the magnetic field equals the thermodynamic 
critical field $H_c$ at which there is bulk two-phase coexistence between the N 
and SC phase. Instead of Eq.(\ref{eq:bulktranstp1}) they obey the asymptotic 
conditions
\begin{equation}
\varphi(\pm \infty)=1 \ \ \ , \ \ \  a(\pm \infty)=0. 
\label{eq:bulktranstp2}
\end{equation}
Remarkably, the calculation of this class of solutions is very simple and
analytic results can be obtained for the entire type-I regime. Indeed, the
asymptotic condition for the vector potential along with the boundary condition 
at $x=0$
imply that $a\equiv 0$ and we are left with a single differential equation for 
$\varphi(x)$ taking the simplified form, for $T<T_c$,
\begin{equation}
\ddot{\varphi}=-\varphi+\varphi^3, \label{eq:gleq1tpmscl}
\end{equation}
with the solution $\varphi(x)=\coth[(x+\delta)/\sqrt 2]$, for $x>0$, and where
$\delta$ is determined by the boundary condition~(\ref{eq:boundtptrans})~\cite{CJB}. This solution must then be combined with a SC/N interface in the limit of $|x| \rightarrow \infty$ as necessary to obey the bulk N condition as initially imposed. This combination defines a macroscopic SC layer. For general values of $\kappa$, the SC/N interface, as well as the solutions describing a finite SC sheath, must be determined numerically. At bulk two-phase coexistence the two solutions described above may occur yielding an interface delocalization transition (or a wetting transition). This transition describes the delocalization of the SC/N interface from the TP into the bulk of the material away from the TP. We now address the analysis of this phase transition.

\subsection{Phase diagram at bulk coexistence}
\label{subsec:pdtranstp}

\begin{figure}[p]
\resizebox{\textwidth}{!}{
   \includegraphics{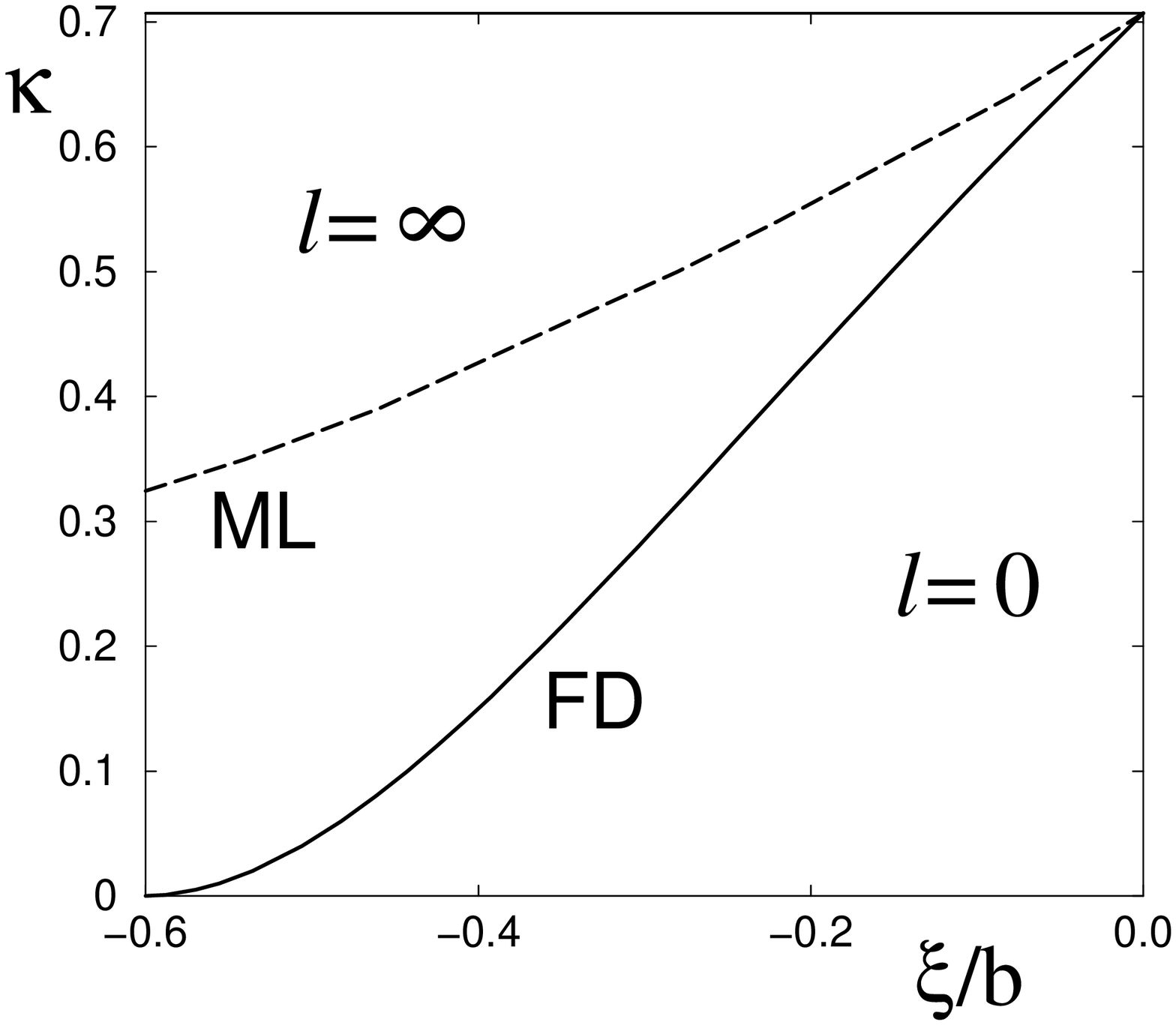}
}   
\caption{Wetting phase diagram at bulk two-phase coexistence for a transparent
twinning plane (TP) in the variables $\kappa$ (with $0 \leq \kappa \leq 1/\sqrt 2$) and $\xi/b$ (with $b<0$).
The solid line  represents the first-order delocalization transition and
separates the region without a SC sheath ($l=0$) from that with a 
macroscopically thick SC layer ($l=\infty$). The dashed line is the 
metastability limit of the null solution corresponding to the normal state of the TP, encountered when, e.g., the temperature is increased towards $T_c$.}
\label{fig:twinfig1}
\end{figure}

In Fig.\ref{fig:twinfig1} we show the phase diagram for a type-I
superconductor with a transparent TP as a function of the parameters $\kappa$ 
and $\xi/b$. The magnetic field $h$ is fixed to its coexistence value 
$h_c=1/\sqrt 2$~\cite{FN1}.
Varying the temperature for a given material, i.e. for fixed $\kappa$, 
corresponds to traveling horizontally in the diagram. Moving along coexistence 
towards $T_c$ corresponds to decreasing $\xi/b$ towards $-\infty$, since 
$\xi\propto|T-T_c|^{-1/2}$ and $b$ is a negative TP constant. 
We distinguish two regions according to the thickness $l$ of the SC sheath near
the TP. In the first region $l=0$ indicating that the stable solution is the
null solution without a SC sheath near the TP, while in the other region, the 
stable solution describes a macroscopic sheath on both sides of the TP and thus 
$l=\infty$. Note that a solution with a strictly finite thickness, i.e. 
$0<l<\infty$, is never stable and plays no significant role. The two regions are separated by the equilibrium surface phase transition line FD (first-order delocalization) while the dashed line ML marks the metastability limit of the N state of the TP, when $T$ is increased towards $T_c$.

To identify the loci of the first-order phase boundary FD we can apply the technique of phase portraits which is a well-known concept in the study of
wetting phase transitions and has proven to be equally instructive for
investigating interfacial phase transitions in superconductors~\cite{IND,IND1,BLO}.
In practice, however, it is simpler to calculate $\gamma$ from 
Eq.~(\ref{eq:gammatp2}) for a macroscopic SC sheath and to see where it equals 
the surface free energy of the null solution. Since $\gamma=0$ in the latter 
state the condition for the delocalization transition reads
\begin{equation}
\gamma_{TP/SC}+\gamma_{SC/N}=0, \label{eq:deloctranstp}
\end{equation}
where $\gamma_{TP/SC}$ is the surface free energy of the bulk SC phase against 
the TP and $\gamma_{SC/N}$ the surface tension of the SC/N interface. The former
can be calculated by considering the first integral of Eq.(\ref{eq:gleq1tpmscl})
(for $h=h_c$)
\begin{equation}
\dot{\varphi}=\frac{1}{\sqrt 2}(1-\varphi^2),
\label{eq:fiint}
\end{equation}
which leads, by inserting into the integral~(\ref{eq:gammatp2}), using also~(\ref{eq:boundtptrans}) and considering the half space $x>0$, to
\begin{equation}
\gamma_{TP/SC}=\frac{2\sqrt 2}{3}-\varphi^2_+\left[\frac{2\sqrt 2}{3}\varphi_+
+\frac{\xi}{b} \right]. \label{eq:gammatpsc}
\end{equation}
Here, $\varphi_+$ is given by the solution of
\begin{equation}
\varphi_+^2+\sqrt 2 \xi\varphi_+/b-1=0.\label{eq:varphi+}
\end{equation}
This equation follows from combining the first integral with the boundary 
condition~(\ref{eq:boundtptrans}). Concerning the surface tension
$\gamma_{SC/N}$ we can use the accurate analytic expression pioneered in~\cite{MIS} and improved in~\cite{BOU},
\begin{equation}
\gamma_{SC/N}=\frac{2\sqrt 2}{3}-1.02817\sqrt \kappa -0.13307\kappa\sqrt\kappa +
{\cal O}(\kappa^2\sqrt\kappa). \label{eq:gammascnanalyt}
\end{equation}
Due to the high accuracy of this expansion when truncated at order
$\kappa\sqrt\kappa$ in the entire type-I regime, we immediately find, by 
inserting~(\ref{eq:gammatpsc}) and~(\ref{eq:gammascnanalyt}) into the 
condition~(\ref{eq:deloctranstp}), an accurate analytic result for the 
first-order phase boundary FD. The deviation from the numerical results 
lies within the thickness of the solid line in Fig.~\ref{fig:twinfig1}, even at 
$\kappa=1/\sqrt 2$. For $\kappa=0$ the transition occurs at $(\xi/b)^*=-0.6022$ 
and expanding the phase boundary about this point reveals that it approaches the $\kappa=0$ axis in a parabolic manner $\kappa(\xi/b)\sim a(\xi/b 
- (\xi/b)^*)^2$, with $a\approx 4.95$. This demonstrates that in the 
low-$\kappa$ regime the system behaves both qualitatively and quantitatively 
precisely like the semi-infinite system with a wall~\cite{IND,IND1,CJB}.

A final aspect of the phase diagram relates to the calculation of the 
metastability limit ML for which it is justified to use the linearized version 
of the GL theory. In this approximation the non-linear terms in the GL 
equations are omitted so that Eq.~(\ref{eq:gleqtp1}) reads, for $T<T_c$,
\begin{equation}
\ddot\varphi_0=-\varphi_0+a_0^2\varphi_0/\kappa^2, \label{eq:gleqtplin}
\end{equation}
while the second one becomes trivial with the general solution
\begin{equation}
a(x)=a_0(x)=h(x+x_0), \label{eq:gleqtp2lin}
\end{equation}
where the boundary condition~(\ref{eq:a0}) immediately gives $x_0=0$. 
Eq.~(\ref{eq:gleqtp2lin}) expresses 
that the magnetic field is at no point expelled by the SC sheath that nucleates 
at ML. Thus we need only solve Eq.~(\ref{eq:gleqtplin}), subject to the boundary conditions
\begin{equation}
\dot{\varphi}_0(0^+)=\frac{\xi}{b}\varphi_0(0^+) \ \ \  , \ \ \  
\varphi_0(\infty)=0.
\label{eq:boundtplin}
\end{equation}
To find $\varphi_0(x)$ we follow Ref.~\cite{IND1} and reduce~(\ref{eq:gleqtplin}) to a first-order (non-linear) differential equation by introducing the function 
$q_0(x)=\dot{\varphi}_0(x)/\varphi_0(x)$ obeying the equation
\begin{equation}
\dot{q}_0+q_0^2=-1+a_0^2/\kappa^2, \label{eq:eqqtp}
\end{equation}
with the boundary condition $q_0(0)=\xi/b$. This equation must be solved such 
that $q_0(x)$ has the acceptable asymptotic behaviour $q_0(x) \sim -h x/\kappa$, implying a Gaussian decay for $\varphi_0(x)$. This is done by performing 
(backwardly) the numerical integration of the auxiliary equation 
\begin{equation}
\dot{q}_0+q_0^2=-1+(hx/\kappa)^2, \label{eq:eqqtp2}
\end{equation}
starting from $q_0(x)=-h x/\kappa$ for large $x$, down to $x=0$. For given $h/\kappa$, the ML value for $\xi/b$ then simply follows from $q_0(0)=\xi/b$. Moreover, using the result for the function $q_0(x)$, we can construct explicitly a solution to~(\ref{eq:gleqtplin}) of the form
\begin{equation}
\varphi_0(x)=\varphi_0(0)\exp \left( \int_0^x q_0(x')dx' \right). \label{eq:solutlintp}
\end{equation}
The line ML is obtained by taking a value of $\kappa$ and applying the 
procedure for the coexistence value $h_c=1/\sqrt 2$. Further, the same method 
applies away from coexistence (as well as to $T>T_c$) where it serves to obtain the critical nucleation field as a function of temperature (see below).

The phase diagram discussed above demonstrates that for all type-I materials 
with an \emph{internal transparent} TP there exists an interface delocalization 
transition at some temperature $T_D$ strictly below $T_c$. This transition can
be interpreted as a genuine wetting transition such that for temperatures 
$T>T_D$ the SC state completely wets the TP and this occurs on both sides of the TP. For a \emph{transparent} TP this phase transition is of \emph{first order} regardless of the value of $\kappa$. This is very different from the
analogous phase diagram for wetting at the surface of the material for which \emph{both first-order and critical transitions} are predicted~\cite{IND,IND1}. 

A TP displaying TPS is characterized by a negative extrapolation length, hence it is not relevant to consider here positive $b$ values. Explicit surface free-energy calculations further reveal that for $b<0$ the SC phase is preferred
by the TP such that $\gamma_{TP/SC}<\gamma_{TP/N}$, with $\gamma_{TP/N}$ the surface free energy of the N phase in bulk, according to~(\ref{eq:bulktranstp1}). This equality is reversed for all $\kappa$ when $\xi/b > 0$ so that the question of wetting by the SC phase is only relevant for $b<0$. In other words, the line of reversal of preferential adsorption in the $(\kappa,\xi/b)$-plane is located at $\xi/b=0$. Also in this respect the wetting phase diagram differs from that for an external surface or wall~\cite{IND1}.

\subsection{Off-of-coexistence phase behaviour}
\label{subsec:offcxtra}

Now we turn our attention to the issue of phenomena outside coexistence which can most easily be clarified by inspecting the magnetic field versus temperature phase diagram for a given material. These diagrams are much more accessible to
experimental verification than the global diagram in Fig.\ref{fig:twinfig1} and
provide a means of comparing our results with the TPS phase diagrams of 
Ref.~\cite{BUZ}.

\begin{figure}[p]
\resizebox{\textwidth}{!}{
   \includegraphics{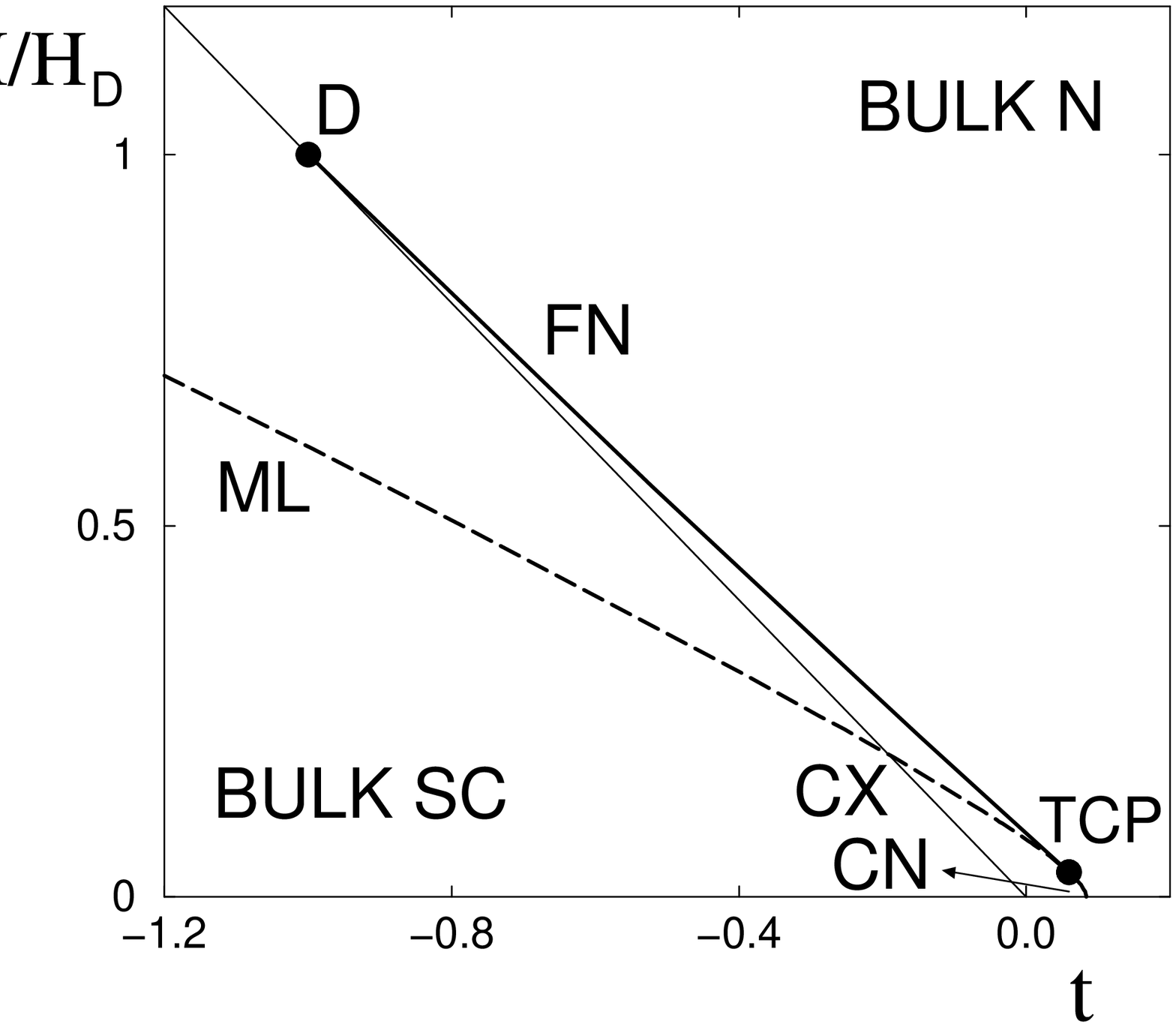}
}   
\caption{Magnetic field versus temperature phase diagram for a transparent TP 
with $\kappa=0.3$ in the variables $H/H_D$ and $t=(T-T_c)/(T_c-T_D)$. At the
nucleation transitions FN and CN local superconductivity appears at the TP.
Further details of the phase boundaries are provided in the text.}
\label{fig:twinfig2}
\end{figure}

Fig.\ref{fig:twinfig2} shows a typical example for $\kappa=0.3$ where we employ units based on the delocalization field $H_D=H_c(T_D)$ and the delocalization 
temperature $T_D$, i.e., we take the ratios $H/H_D$ and $t=(T-T_c)/(T_c-T_D).$
A first ingredient in this diagram is the bulk coexistence line CX with two anchor points, the delocalization point D at $H/H_D=1$ and $t=-1$ and the bulk critical point at $H/H_D=0$ and $t=0$. Since the delocalization transition is first-order the point D is the starting point for a \emph{prewetting} line FN (first-order nucleation) which marks a first-order transition between the null solution with $l=0$ and a finite sheath. The line FN is tangential to the coexistence line CX at D and changes into a continuous or critical nucleation line CN at a tricritical point TCP. The short stretch CN can be computed using the technique discussed earlier to determine the line ML in
Fig.~\ref{fig:twinfig1}. It is important to stress that this transition is
critical since the sheath will appear with an infinitesimal amplitude. 
Concerning the spatial extension in the direction perpendicular to the TP, 
however, it turns out that the sheath always has a thickness of the order of 
$\xi$ even just below the transition. Note that at the tricritical point FN and CN meet with common tangents and, to the left of TCP, CN continues as the metastability limit ML of the N state of the TP (dashed line). Finally, the line CN ends at zero field at the temperature $T_{{\rm c,TP}}$ which obeys the equation~\cite{IND,IND1}
\begin{equation}
\frac{\xi(T_{{\rm c,TP}})}{b}=-1, \label{eq:tptcsb}
\end{equation}
showing that the extrapolation length $b$ can be found in principle from 
experimental determination of $T_{{\rm c,TP}}$.

We obtain the first-order nucleation line FN by numerically solving the GL 
equations for a sheath-type solution and finding the point in the phase diagram
at which the free energy of this solution is zero. The location of TCP can be 
determined accurately by extending the zeroth-order solution of the linear GL 
equations (see above) a little further. In order to obtain a non-vanishing solution 
of the non-linear theory we need a correction to $a_0(x)$ and $\varphi_0(x)$. In view of Eq.~(\ref{eq:solutlintp}) $\varphi_0(x)$ is the small quantity through 
its small amplitude and thus it suffices to expand $a(x)=a_0(x)+a_1(x)+\ldots.$
Using the first integral of the equations and working to second order in 
$\varphi_0$ we immediately get
\begin{equation}
\dot a_1= \frac{1}{2h}(\frac{a_0^2\varphi_0^2}{\kappa^2}- 
\dot \varphi_0^2 \pm \varphi_0^2),
\label{eq:correcatp}
\end{equation}
or, using (\ref{eq:gleqtplin}),
\begin{equation}
\dot a_1= \frac{1}{2h}\dot q_0(x) \varphi_0^2(x). \label{eq:correcatp2}
\end{equation}
This result is particularly useful when we use it in an alternative expression for $\gamma$ evaluated in the extrema (see~\cite{IND1})
\begin{equation}
\gamma=\int_{-\infty}^{+\infty} {\rm d}x \left\{-\frac{\varphi^4}{2}+(\dot a -h)
^2\right\}. \label{eq:alterntp1}
\end{equation}
The advantage of this expression is twofold, firstly the boundary term is 
absorbed and secondly it shows that both terms are small, of order $\varphi_0^4$
for small $\varphi_0$. Along the line CN, $\gamma=0$ because $\varphi_0=0$ and 
$\dot a_0=h$. Another mechanism, however, to make $\gamma$ vanish is the compensation of the two terms in~(\ref{eq:alterntp1}), which can be put in the form
\begin{equation}
2h^2=\int_{-\infty}^{+\infty}{\rm d}x \dot q_0^2(x) \varphi_0^4(x) /
\int_{-\infty}^{+\infty}{\rm d}x \varphi_0^4(x). \label{eq:tricritp}
\end{equation}
and which is of use in determining TCP as follows. For each value of $\kappa$ a
continuous set of $(h,\xi/b)$ pairs can be found from the linear theory, but only one pair will satisfy the condition~(\ref{eq:tricritp}) for the onset of the nucleation of a finite (in contrast with infinitesimal) sheath, which is then by definition the TCP. 

The nucleation transitions discussed above indicate the point at which local 
superconductivity sets in near the TP, which obviously corresponds to the 
phenomenon of TPS. Consequently the results presented here should agree with the predicted TPS phase diagrams for a parallel magnetic field, at least with those 
pertaining to strongly transparent TP's~\cite{BUZ}. We have checked this for 
various $\kappa$ values and indeed found an excellent agreement for all the 
transitions. Furthermore, we may interpret the TPS transition as a genuine 
\emph{prewetting transition} and thus by reinterpreting the existing theoretical and experimental TPS phase diagrams we stress that prewetting has since long been observed in classical superconductors. Intriguingly, we recall that only in
more recent years has clear evidence of prewetting transitions been found in 
experiments on classical binary liquid mixtures~\cite{KEL}. An important feature of the prewetting line found here is that it does not end at a surface critical point as typically predicted and found in liquid mixtures~\cite{DIE,KEL} but becomes, via a tricritical point, a continuous nucleation line. Further, our results are particularly interesting with respect to the behaviour near the wetting point D, since from the experimental TPS diagram it is unclear what happens to the TPS transition near the bulk critical field $H_c$. We now understand that the surface phase transition at the TP becomes a bulk transition precisely at the delocalization transition.

\section{Opaque twinning planes}
\label{sec:opaque}

\subsection{Classification of solutions and terminology}
\label{subsec:classol}

In this section we focus on the second limiting case, i.e., that of completely 
\emph{opaque} TP's for which $\alpha_{{\rm TP}} \rightarrow \infty$. In this case the two sides of the plane are, to some extent, independent since the requirement of a continuous order parameter at $x=0$ is no longer applicable, hence for a general solution $\varphi_+\neq\varphi_-$. However, there remains some coupling between the two half-spaces due to the vector potential and the magnetic induction which must both be continuous at the TP. Moreover, the number of possible solutions is further restricted by the assumption that the extrapolation length $b$ is equal on both sides of the TP. This means that the critical temperature $T_{{\rm c,TP}}$ (which is defined for zero magnetic field) is assumed to be the same for both sides. Nevertheless, a large number of different solutions can still be found among which several are characteristic for the opaque limit and deserve special attention here. For a neatly arranged description of the phases and phase transitions it is appropriate at this stage to present a classification and a schematic overview of the various solutions at bulk coexistence with equal bulk (N) conditions on the two sides of the TP.

\begin{figure}[p]
\resizebox{\textwidth}{!}{
   \includegraphics{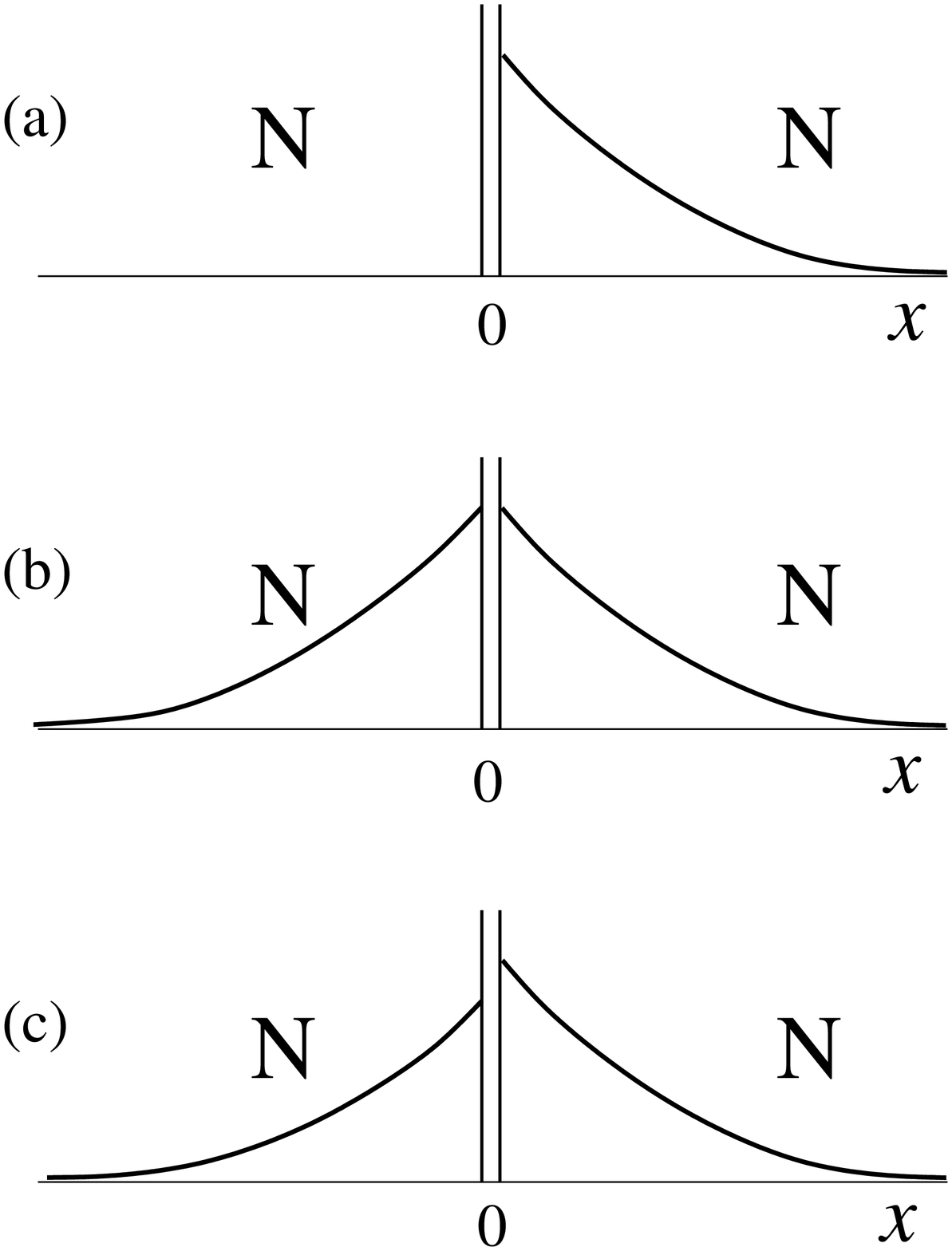}
}   
\caption{Sketch of the order parameter profile for sheath-type
solutions: (a) Wall solution with $\varphi(x)=0$ for $x<0$ and $\dot a (0)=h$.
(b) Continuous solution with $\varphi_+=\varphi_-$ and $a(0)=0$. (c)
Discontinuous solution with $\varphi_+ \neq \varphi_-$ and $a$ and $\dot a$
continuous at $x=0$.}
\label{fig:twinfig3}
\end{figure}

\begin{figure}[p]
\resizebox{\textwidth}{!}{
   \includegraphics{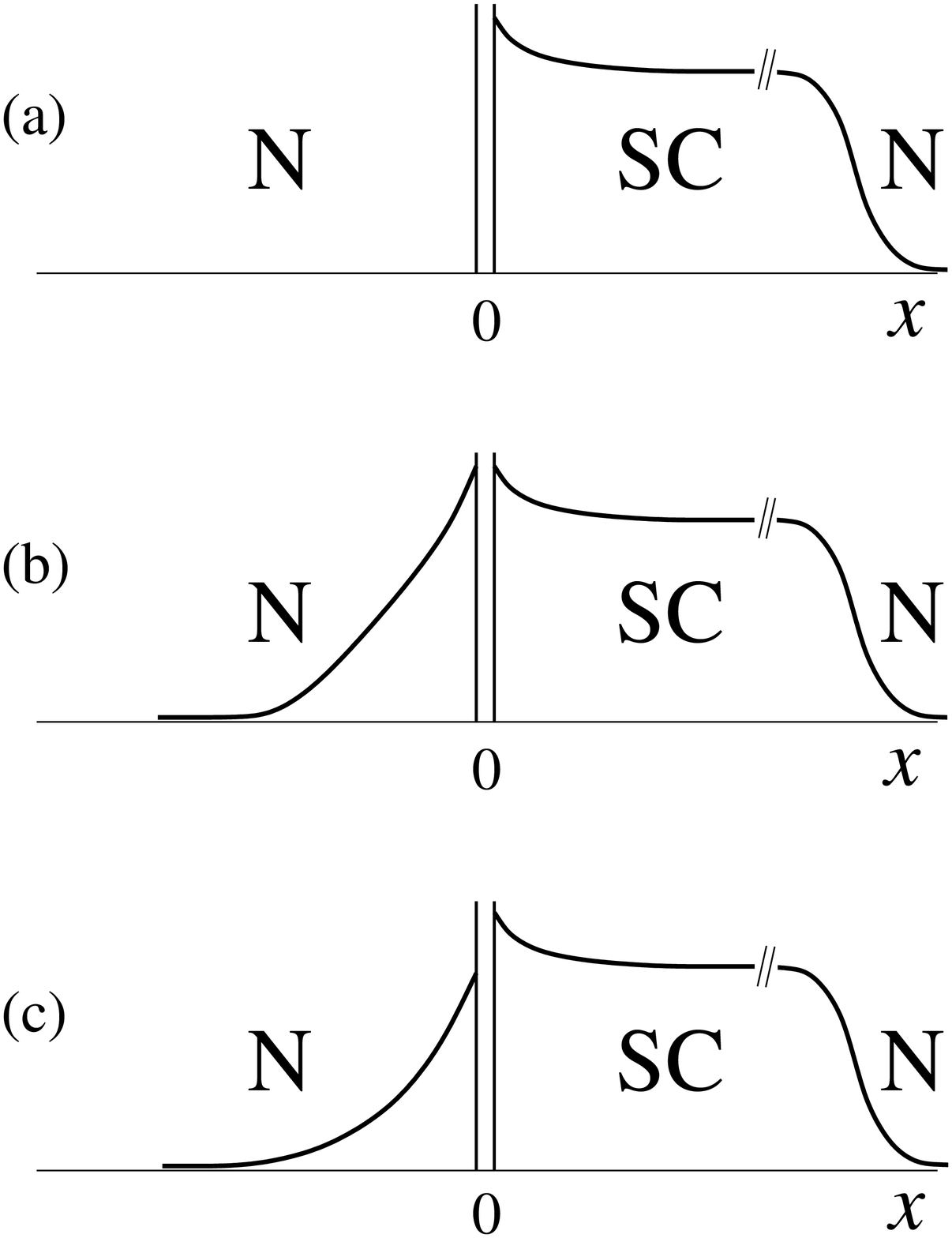}
}   
\caption{Sketch of the order parameter profile for solutions with one
macroscopically thick SC layer: (a) Wall solution with $\varphi(x)=0$ for $x<0$.
(b) Continuous solution with $\varphi_+=\varphi_-$ and a finite sheath for
$x<0$.(c) Discontinuous solution with $\varphi_+ \neq \varphi_-$ and as in (b)
there is a finite sheath for $x<0$.}
\label{fig:twinfig4}
\end{figure}

\begin{figure}[p]
\resizebox{\textwidth}{!}{
   \includegraphics{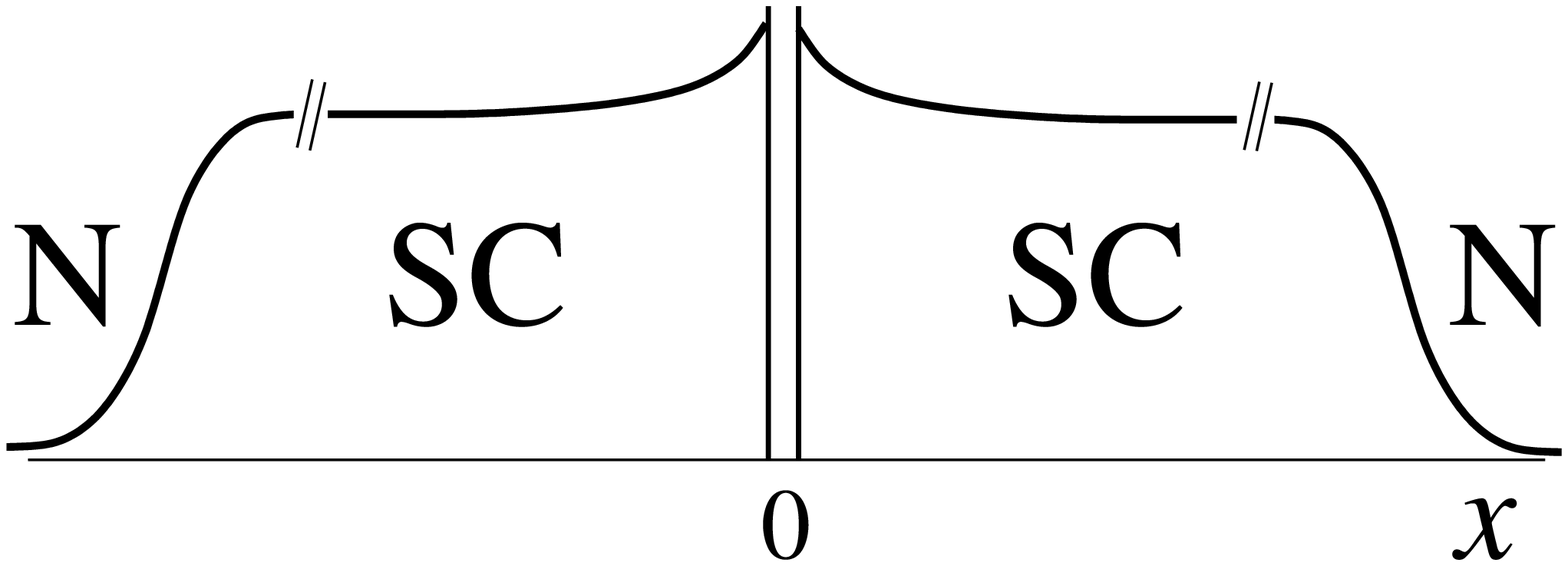}
}   
\caption{Schematic order parameter profile for a continuous solution with a
double macroscopic superconducting layer. Both sides of the TP are in this case
wetted by the SC phase.}
\label{fig:twinfig5}
\end{figure}

A first (trivial) solution is the null solution $\varphi(x)=0$ without local
superconductivity. For solutions with at least one finite SC sheath we 
distinguish three scenarios as depicted in Fig.~\ref{fig:twinfig3}. 
Interestingly, due to the presumed opacity we can consider imposing $\varphi(x)\equiv 0$ for one half-space, say $x<0$, leading to a solution as shown in Fig.~\ref{fig:twinfig3}(a) and referred to as ``wall solution" since it corresponds exactly to the solution found in a system with a wall obeying the boundary condition $\dot a(0)=h$. In the case of a double SC sheath we can have either a continuous solution with $\varphi_+=\varphi_-$ and $a(0)=0$ (Fig.~\ref{fig:twinfig3}(b)) corresponding to the sheath solution for a 
transparent TP, or a more general discontinuous solution with $\varphi_+\neq\varphi_-$ and $a$ and $\dot a$ continuous at $x=0$ (Fig.~\ref{fig:twinfig3}(c)). Arguing along the same lines we also find a series of solutions with a macroscopic SC layer on only one or on both sides of the TP. For the former, in the other half-space, we can either have the wall solution with 
$\varphi(x)\equiv 0$ or a finite sheath with a continuous or discontinuous order parameter at the TP. The order parameter profiles for these solutions are schematically drawn in
Fig.~\ref{fig:twinfig4}. 
For a double macroscopic layer, on the other hand, we should in principle take 
into account both the continuous and discontinuous solution. It
can be shown, however, by a simple argument (see below) that the latter is impossible and therefore unnecessary to consider. As outlined in the previous section a double macroscopic layer will completely expel the magnetic field such that $a=0$ over the entire region of the SC layer. Consequently expression~(\ref{eq:fiint}) applies also here and at $x=0^{+,-}$ this yields 
\begin{equation}
\dot\varphi_{+,-}=\frac{1}{\sqrt 2}(1-\varphi^2_{+,-}), \label{eq:fasepoptp1}
\end{equation}
While it may appear that any combination of $\varphi_+$ and 
$\varphi_-$ obeying these equations is a possible solution, this is only true if
the boundary conditions at the TP are omitted. Indeed, by applying the boundary
conditions which in this case are again given by Eq.~(\ref{eq:boundtptrans}) it is straightforward to see that, since $b$ is assumed to take the same value on both sides of the TP, $\varphi_+$ must equal $\varphi_-$ for a solution with two macroscopic SC layers as exemplified in Fig.~\ref{fig:twinfig5}. Note that this solution is the one we found for the case of a transparent TP.

From the above classification of solutions it is obvious that an opaque TP is in
a sense a combination of a system with a wall and one with a transparent TP, a 
feature which gives rise to additional phase transitions. Before further 
embarking on the phase behaviour it is convenient to comment here briefly on the
terminology used below. The transition from the null solution to a state with 
either one or two finite sheaths as well as the transition from a single sheath
to a pair of sheaths is referred to as a \emph{nucleation} phenomenon. Likewise, the transition from a ``wall state" with one macroscopic SC layer to a state that is the combination of such a layer and a finite sheath is called \emph{nucleation}. Further, the appearance of one macroscopic SC layer, either from the null solution or from one finite sheath, is a \emph{delocalization} or \emph{wetting transition}, while a \emph{depinning transition} refers to the 
transition between one and two macroscopic layers. The latter describes the 
depinning of a SC/N interface that is initially pinned at the TP and is analogous to the one investigated in Ref.~\cite{BAC}. We start with a discussion of the TP states at bulk two-phase coexistence.

\subsection{Phase diagram at bulk coexistence}
\label{subsec:pdopaque}

\begin{figure}[p]
\resizebox{\textwidth}{!}{
   \includegraphics{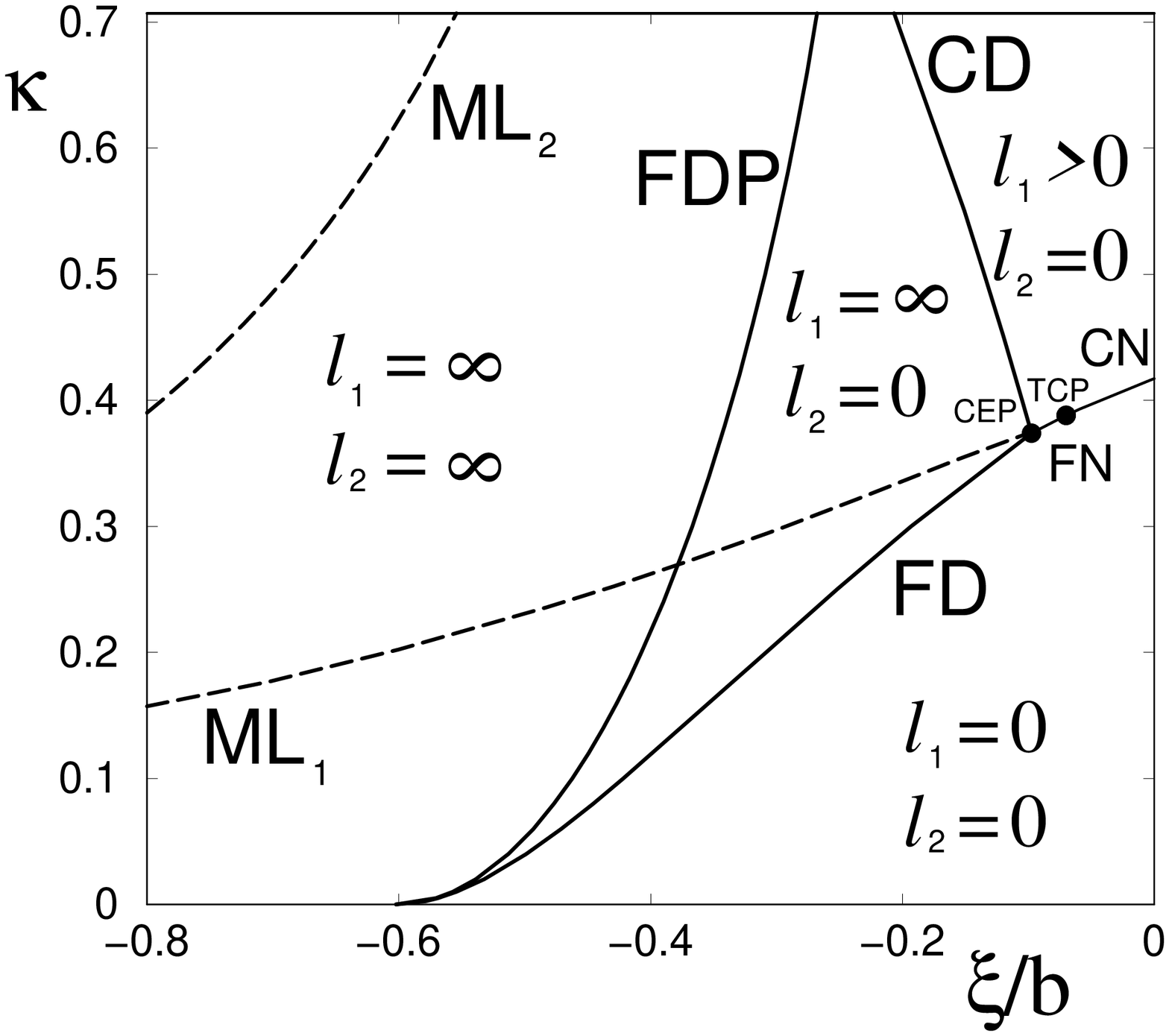}
}   
\caption{Wetting phase diagram at bulk two-phase coexistence for an opaque
TP in the variables $\kappa$ and $\xi/b$. The different phases are characterized
by the values $l_1$ and $l_2$ of the thicknesses of the SC sheaths on the two
sides of the TP. The various transitions and transition lines are explained in
the main text.}
\label{fig:twinfig6}
\end{figure}

The fundamental diagram for interfacial phase transitions in the case of opaque 
TP's is presented in Fig.~\ref{fig:twinfig6}.
The different regions are distinguished according to the values of the two 
length scales $l_1$ and $l_2$ representing the thicknesses of the sheaths on the
two sides of the TP. We assume here, without loss of generality, that $l_1 \geq l_2$. It is striking how different this phase diagram is compared to that for the case of perfect transparency, see Fig.~\ref{fig:twinfig1}. As we will demonstrate below, the main part of the diagram for the opaque case simply recovers the results of the system with a wall~\cite{IND}. It is important to stress at this stage that we are only interested in the wall results for negative $b$ values since this is the relevant region for TP's that show TPS. 

One ingredient of Fig.~\ref{fig:twinfig6} concerns the stability of the 
sheath solutions given in Fig.~\ref{fig:twinfig3}. From our analysis in 
Section~\ref{sec:transp} we know that at bulk coexistence the double symmetric 
sheath (Fig.~\ref{fig:twinfig3}(b)) is never stable, and hence plays no 
significant role. Furthermore, explicit free energy calculations reveal that the discontinuous double sheath (Fig.~\ref{fig:twinfig3}(c)) has an even higher 
free energy than the continuous one and thus is irrelevant. The single sheath 
solution (Fig.~\ref{fig:twinfig3}(a)) is stable for a certain interval of 
$\xi/b$ values provided that $\kappa>0.374$ as can be inferred from the results 
for a system with a wall~\cite{IND,IND1}. Hence, in the upper right corner of 
the diagram we retrieve a feature of the wall diagram, with a finite sheath on 
only one side of the TP while the other side is still in the N phase, thus $l_1>0$ and $l_2=0$. In the lower half of the diagram we have on the low-$\xi/|b|$ side a region where no stable sheath solution exists 
other than the null solution ($l_1=0$ and $l_2=0$). The two regions are 
separated either by the line CN (critical nucleation) ending in a 
tricritical point TCP, or by the short stretch FN (first-order 
nucleation) between TCP and a critical end point CEP. To the left of 
TCP the line CN continues as the metastability limit ${\rm 
ML_1}$ of the null solution.

Next we need to consider the solutions of Fig.~\ref{fig:twinfig4}
and~\ref{fig:twinfig5} and to investigate their stability. Our calculations
suggest that when the temperature is increased towards $T_c$ the system first 
enters the state given in Fig.~\ref{fig:twinfig4}(a), either from one finite 
sheath or from the null solution. These transitions are indicated by the line 
CD (critical delocalization) and FD (first-order delocalization) respectively, 
meeting each other at the CEP. By further increasing the temperature the system 
undergoes a \emph{first-order depinning transition} towards the double 
symmetric solution, i.e. a transition from the solution given in 
Fig.~\ref{fig:twinfig4}(a) towards the solution of Fig.~\ref{fig:twinfig5}. Thus the regions $l_1=\infty$, $l_2=0$ and $l_1=l_2=\infty$ are separated by the 
first-order phase boundary FDP (first-order depinning). Lastly, the line 
${\rm ML_2}$ represents the metastability limit of the state with $l_1=\infty$, $l_2=0$, with the SC/N interface pinned at the TP. Note that the solutions represented in Fig.~\ref{fig:twinfig4}(b) and~(c) are never stable.

An important conclusion that we draw from this phase diagram is that for relatively low temperatures an opaque TP behaves as a system with a wall, with 
on one side of the TP the different wall solutions while the other side remains 
fully in the N phase. In other words, to the right of the line FDP in 
Fig.~\ref{fig:twinfig6} we immediately obtain the phase diagram by copying the 
results for a wall without additional computations. All the details concerning
the determination of the transition lines for this part of the diagram can be 
found in Ref.~\cite{IND1} and~\cite{CJB}. In particular, the critical delocalization condition reads
\begin{equation}
\gamma_{W/N}=\gamma_{W/SC}+\gamma_{SC/N},
\label{eq:critopaqtp}
\end{equation}
while the first-order delocalization condition is simply given by
\begin{equation}
\gamma_{W/SC}+\gamma_{SC/N}=0, 
\label{eq:foopaqtp}
\end{equation}
where $\gamma_{W/N}$ ($\gamma_{W/SC}$) is the surface free energy of the bulk N
(SC) phase against a wall.
The novel feature in Fig.~\ref{fig:twinfig6} is the FDP line arising 
from the possibility of having a macroscopic SC layer also in the other region leading to a depinning transition at higher temperatures along with the associated spinodal ${\rm ML_2}$. It is precisely these two transition lines that we will be concentrating on in the remainder of this section.

The condition for the first-order depinning transition is obtained by equating 
the free energy of the solution with one macroscopic layer at $x>0$ and no sheath at $x<0$ to that of the solution with the double symmetric macroscopic SC layer. The former can be written conveniently as the sum $\gamma_{W/SC}+\gamma_{SC/N}$, thus the depinning condition reads
\begin{equation}
\gamma_{W/SC}+\gamma_{SC/N}=2(\gamma_{TP/SC}+\gamma_{SC/N}). 
\label{eq:depinopaqtp}
\end{equation}
In general the depinning phase boundary has to be calculated numerically. 
An approximate analytic result can be obtained using the powerful expansions in 
$\kappa$ for the surface free energies. For the surface tension $\gamma_{SC/N}$ 
we use the result~(\ref{eq:gammascnanalyt}) while a similar expansion has been 
derived for $\gamma_{W/SC}$ in Ref.~\cite{CJB}. Using these earlier results in 
the depinning condition~(\ref{eq:depinopaqtp}) provides a very accurate 
approximation for the phase boundary across the complete type-I range with an 
error less than $1\%$. For $\kappa=0$ the depinning and the delocalization 
transitions coincide at $(\xi/b)^*=-0.6022$. Both phase boundaries have a 
parabolic foot near $\kappa=0$, i.e. $\kappa(\xi/b)\sim a(\xi/b-(\xi/b)^*)^2$ 
with $a\approx 27.0$ for depinning and $a\approx 4.95$ for
delocalization~\cite{CJB}. From a mathematical point of view it is interesting to examine whether the extensions of the lines FDP and CD meet in the type-II regime, for $\kappa > 1/\sqrt{2}$. The condition for a line crossing is given by combining~(\ref{eq:critopaqtp}) and~(\ref{eq:depinopaqtp}). Explicit calculations using an expansion in $\kappa$ for the line CD obtained in~\cite{VLHA} reveal indeed an intersection at $\kappa \approx 0.815$, $\xi/b \approx -0.251$.

The line ${\rm ML_2}$ marks the nucleation of an infinitesimal sheath on one side of the TP under the condition that on the other 
side of the TP a macroscopic sheath is stable. To obtain this line we first
compute numerically a solution for a macroscopic layer in one half space, $x>0$ 
say, for a given value of $\xi/b$ subject to the condition $\dot a(0)=h$. To
proceed we combine it with a solution of the linear GL theory applied to the
region $x<0$ using the scheme introduced in Section~\ref{sec:transp}. The 
zeroth-order solution for the vector potential is again given by 
Eq.~(\ref{eq:gleqtp2lin}) with in this case $x_0$ determined by the value of 
$a(0)$ which we get from the numerical calculation in $x>0$. Finally we have to 
compare $q_0(x_0)$ with the given value of $\xi/b$ and iterate the procedure 
such that the two become equal. This defines the critical nucleation of the 
sheath in the region $x<0$. The line ${\rm ML_2}$ is now formed by applying this method for the coexistence field $h_c$. The procedure is equally applicable for 
any other field appropriate for the study of off-of-coexistence phenomena, an 
issue which is addressed below.

\subsection{TPS phase diagrams}
\label{subsec:pdopaque2}

From the properties of an opaque TP at bulk coexistence elucidated above we 
anticipate that the onset of local superconductivity near this plane generally 
happens in two steps. This is a natural consequence of the stability of ``wall 
solutions" in these systems with $\varphi(x)=0$ on one side of the TP while on the other side the SC phase has already nucleated. We now wish to concentrate on the various nucleation phenomena, the order of which depends on the parameter $\kappa$, as a function of magnetic field and temperature.

\begin{figure}[p]
\resizebox{\textwidth}{!}{
   \includegraphics{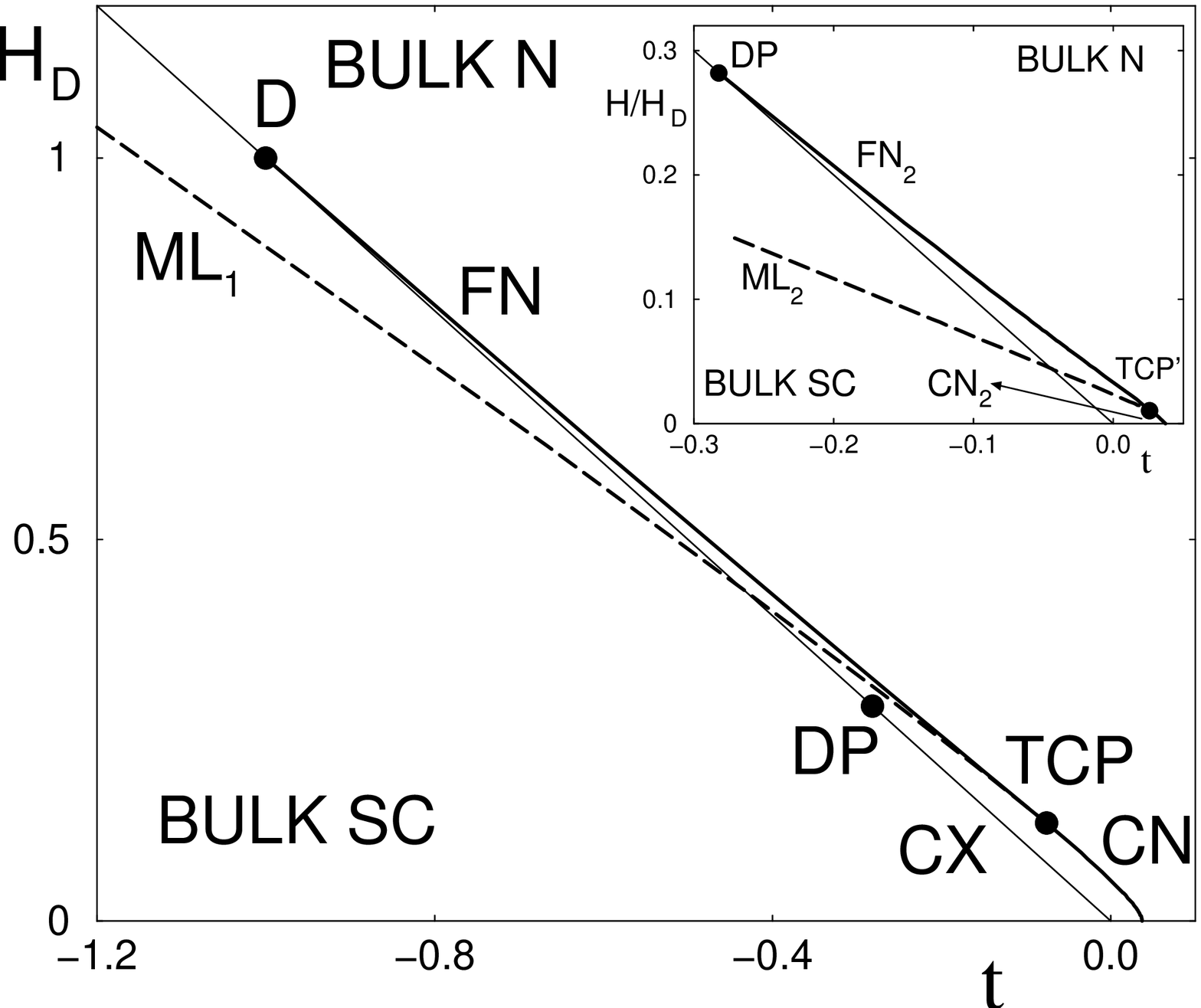}
}   
\caption{Surface phase diagram for an opaque TP with $\kappa=0.3$ in the
variables $H/H_D$ and $t=(T-T_c)/(T_c-T_D)$. The inset shows the different
nucleation transitions near the first-order depinning point DP. All the
transitions are explained in the text.}
\label{fig:twinfig7}
\end{figure}

For $\kappa<0.374$ the delocalization transition on one side of the 
TP is first-order and thus will have a prewetting extension into the region of 
the phase diagram where the N phase is stable in bulk. This prewetting line 
FN is the first-order nucleation of a sheath on one side of the TP, and
hence corresponds to the transition between the null solution and a solution as 
given in Fig.~\ref{fig:twinfig3}(a). This line will be tangential to the bulk coexistence line CX at the first-order delocalization point D and changes at the tricritical point TCP into a continuous nucleation line CN. The latter describes the nucleation of one infinitesimal sheath and continues as the metastability limit ${\rm ML_1}$ of the null solution to the left of TCP. So far the surface phase diagram, an example of which is given in Fig.~\ref{fig:twinfig7} for 
$\kappa=0.3$, resembles the diagram for the system with a wall~\cite{IND,IND1}. Note that we use the same units as in Fig.~\ref{fig:twinfig2}. 

As we have argued above, however, another distinct point exists at bulk 
coexistence at higher temperatures but still below $T_c$, namely the first-order depinning point DP. The presence of the depinning transition leads to a variety 
of additional transitions compared to the wall system. Indeed, due to the 
first-order character of this point a second first-order nucleation line ${\rm
FN}_2$ appears, attached to the point DP. This phase boundary can be interpreted as a \emph{predepinning} line. For the sake of clarity this line is omitted in the main figure, with all the details near the depinning transition given in the inset of Fig.~\ref{fig:twinfig7}. At the line ${\rm FN}_2$ the SC phase nucleates on the other side of the TP, hence it is the transition between 
one sheath and a double sheath. The line ${\rm FN}_2$ is tangential to the line CX at DP and meets a continuous nucleation line ${\rm CN}_2$ at a second tricritical point ${\rm TCP'}$. To the left of ${\rm TCP'}$, ${\rm CN}_2$ continues as the metastability limit ${\rm ML_2}$ of the state with one sheath. Obviously ${\rm CN}_2$ denotes the critical nucleation of the second SC sheath. The two critical nucleation lines end in zero field at the
same point $t_{{\rm c,tp}}$ which follows from the assumption that the TP is
characterized by a single value of the parameter $b$.

For materials with $\kappa >0.374$ for which the delocalization transition is critical, the field-temperature phase diagram will undergo qualitative changes as regards the wetting-like transitions, i.e., the transitions related to the nucleation of the SC phase on \emph{one} side of the TP. We refer interested readers to Ref.~\cite{IND1} for examples of $(H,T)$-phase diagrams containing these nucleation transitions and restrict ourselves here to a brief discussion of the pertinent features relevant for the present study. For $\kappa=0.374$ the point D and the starting point of the first-order nucleation line FN separate and by further increasing $\kappa$ this first-order transition disappears. The critical nucleation line CN then extends to low temperatures without intersection with the bulk coexistence line CX. Hence in this case the nucleation of the first sheath is always critical. For the nucleation of the second sheath there is still the possibility of first-order as well as second-order transitions. In fact, we know that the depinning transition is first-order for all type-I materials so that the situation that we sketched for $\kappa=0.3$ near the depinning point DP is representative for the entire type-I regime. Thus for an \emph{opaque} TP there are always two distinct nucleation transitions, i.e. the SC phase never appears simultaneously on both sides of the TP.

\section{Discussion and concluding remarks}
\label{sec:conc}

In this paper we have analyzed the phase behaviour of the SC/N interface near 
internal TP's in type-I superconductors. Our calculations reveal that the 
results are highly sensitive to the boundary conditions imposed at the TP, which in turn depend on the degree of transparency of the TP for electrons. For 
perfectly transparent TP's the order parameter is continuous such that only 
symmetric profiles need to be considered and at first sight the analysis resembles that of an external surface or wall. The only differences originate from the boundary condition for the vector potential and, surprisingly, this small technical modification turns out to have dramatic consequences for the order of the delocalization transition. In particular, \emph{for a transparent TP only first-order transitions are found for the entire type-I regime} while the possibility of critical transitions that are predicted for the wall system~\cite{IND,IND1} are suppressed. Our results for the magnetic
field versus temperature phase diagrams are in excellent agreement with earlier 
experimental and theoretical results obtained in the context of TPS~\cite{BUZ}.

We comment further that our approach for the \emph{nucleation} of the SC phase near the TP equally applies to type-II materials (with $\kappa>1/\sqrt 2$). Specifically, we have considered the case of Nb with $\kappa\approx 1$ for which the experimental results suggest that the properties of the TP are very close to the limiting case of perfect transparency. The nucleation transition in this case is always of second-order and our results again perfectly agree with earlier work~\cite{BUZ}. We remark that in this case the nucleation lines are only relevant at sufficiently high fields above their intersection with the upper critical field $H_{c2}$.

The situation drastically changes when considering opaque TP's in which case a 
discontinuity in the order parameter is allowed at the TP. The decoupling of the two sides of the TP makes it possible to consider wall solutions with $\varphi(x)=0$ on one side of the TP. Moreover, from free-energy considerations, we have shown that these solutions are stable in a large region of the phase space and, as a result, the system precisely undergoes the transitions predicted for a wall system. In this case only one side of the TP will be wetted by the SC phase at the delocalization transition which can be either first-order or critical depending on the value of $\kappa$. By further increasing the temperature at bulk coexistence a first-order depinning transition is predicted for all type-I materials. Consequently, the field-temperature phase diagrams for opaque TP's fundamentally differ from their counterparts for the transparent limit. A characteristic feature of an opaque TP is the existence of two distinct nucleation lines which in principle should be measurable and thus can \emph{provide an experimental means of distinguishing between transparent and opaque TP's}. Related to this we remark that experiments in Sn ($\kappa \approx 0.13$) have revealed only one nucleation transition, although it is assumed that the twin boundary in this material has a low transmission for electrons~\cite{BUZ,MIN,SAM}. The apparent absence of a second nucleation transition can in this case be attributed to the low-$\kappa$ value of the material, since in the low-$\kappa$ regime the various transition lines lie extremely close together (see Fig.~\ref{fig:twinfig6}) and they would be difficult to distinguish in an experiment. Note that in the limit 
$\kappa\rightarrow0$ the differences between a transparent TP, an opaque TP and 
an external surface vanish. 

Finally, for type-II materials with opaque TP's our calculations show that the nucleation can be either first-order or critical and this is at variance with the conclusion of earlier work~\cite{AVE} stating that the TPS transition is always second-order for type-II materials. Our study demonstrates that this surmise is correct only for the case of transparent TP's. The reason for this is that the tricritical nucleation point TCP merges with the delocalization transition at $\kappa=1/\sqrt 2$ for transparent TP's. This can be seen also from the merging of the spinodal ML with the line FD in Fig.~\ref{fig:twinfig1}. In contrast, for opaque TP's the TCP of nucleation remains well off of coexistence, at $H>H_c$, and in the type-II regime at $H > H_{c2}$, so that there is room for first-order nucleation. This is further demonstrated by the fact that, in Fig.~\ref{fig:twinfig6}, the lines ${\rm ML_2}$ and FD are still far apart at $\kappa=1/\sqrt 2$.

One remarkable implication of our work is that, in general, with the exception of perfectly transparent TP's, there exist stable states of local superconductivity which are \emph{asymmetric} about the TP. The possibility that a SC sheath or a macroscopic SC layer can occur on one side of the TP while the other side is in the normal state is indeed noteworthy and has been met with scepticism. It has been suggested that, since our analysis is essentially one-dimensional and neglects states which are inhomogeneous in the direction(s) parallel to the TP, there may exist modulated states , e.g., composed of a linear array of soft vortices parallel to the TP~\cite{SAM} or states with a local field penetration and a change of phase of the wave function, which may have a lower free energy than the states we have considered~\cite{BUZP}. Although we cannot rule out this possibility at present, we would like to point out that asymmetric wetting, followed by symmetric depinning, has been found previously in the context of grain-boundary wetting~\cite{EBHA}, in the framework of a real scalar order parameter model of Ising type.

\section*{Acknowledgements}

We thank Chris Boulter, Todor Mishonov and Alexander Buzdin for stimulating discussions. This research has been supported by the Belgian Fund for Scientific Research (FWO-Vlaanderen), the Inter-University Attraction Poles (IUAP) and the Concerted Action Research Programme (GOA). Most of the results of this paper have been obtained in the framework of two theses~\cite{FC}.


\begin{thebibliography}{99}
\bibitem{FN1} This value for the thermodynamic field $h_c$ is easily
inferred from the functional for $\gamma$, Eq.(\ref{eq:gammatp2}). Indeed, the free energy density approaches in the bulk SC state the value $-1/2+h^2$ and thus $\gamma$ would diverge unless $h=h_c=1/\sqrt 2$.
\end{thebibliography}
\end{document}